\documentclass[aps,prr,superscriptaddress, twocolumn]{revtex4-1}

\usepackage[english]{babel}
\usepackage{times}
\usepackage{graphicx}
\usepackage{graphics}
\usepackage{amsmath}
\usepackage{mathtools}
\usepackage{amsfonts}
\usepackage{amssymb}
\usepackage{dsfont}
\usepackage{epstopdf}
\usepackage{makeidx}
\usepackage{subfigure}
\usepackage{color}
\usepackage{pgf}
\usepackage{bm}
\usepackage{ulem}
\usepackage{tikz} 
\begin{document}

\title{Kuramoto synchronization of quantum tunneling polarons for describing the dynamic structure in cuprate superconductors}

\author{Victor Velasco} 

\affiliation{Instituto de Fisica, Universidade Federal do Rio de Janeiro, Caixa Postal
68528, Rio de Janeiro, Brazil}

\author{Marcello B. Silva Neto} 

\affiliation{Instituto de Fisica, Universidade Federal do Rio de Janeiro, Caixa Postal
68528, Rio de Janeiro, Brazil}

\author{Andrea Perali}

\affiliation{School of Pharmacy, Physics Unit, Università di Camerino, Via Madonna delle Carceri 9, 62032 Camerino, Italy}

\author{Sandro Wimberger}
\affiliation{Dipartimento di Scienze Matematiche, Fisiche e Informatiche, Università di Parma, 43124 Parma, Italy}
\affiliation{INFN, Sezione di Milano Bicocca, Gruppo Collegato di Parma, 43124 Parma, Italy}

\author{Alan R. Bishop}
\affiliation{Center for Nonlinear Studies, Los Alamos National Laboratory, Los Alamos, NM 87545, U.S.A.}

\author{Steven D. Conradson}
\affiliation{Department of Complex Matter, Josef Stefan Institute, 1000 Ljubljana, Slovenia}
\affiliation{Department of Chemistry, Washington State University, Pullman, WA 90164, U.S.A.}

\date{\today} 

\begin{abstract}
A major open topic in cuprates is the interplay between the lattice and electronic dynamics and the importance of their coupling to the mechanism of high-temperature superconductivity (HTSC). As evidenced by Extended X-ray Absorption Fine Structure experiments (EXAFS), anharmonic structural effects are correlated with the charge dynamics and the transition to a superconducting phase in different HTSC compounds. Here we describe how structural anharmonic effects can be coupled to electronic and lattice dynamics in cuprate systems by performing the exact diagonalization of a prototype anharmonic many-body Hamiltonian on a relevant $\mathrm{6-}$atom cluster and show that the EXAFS results can be understood as a Kuramoto synchronization transition between coupled internal quantum tunneling of polarons associated with the two-site distribution of the copper-apical-oxygen ($\mathrm{Cu-O_{ap}}$) pair in the dynamic structure. The transition is driven by the anharmonicity of the lattice vibrations and promotes the pumping of charge, initially stored at the apical oxygen reservoirs, into the copper-oxide plane. Simultaneously, a finite projection of the internal quantum tunneling polaron extends to the copper-planar-oxygen ($\mathrm{Cu-O_{pl}}$) pair. All these findings allow an interpretation based on an effective quantum mechanical triple-well potential associated with the oxygen sites of the 6-atom cluster, which accurately represents the phase synchronization of apical oxygens and lattice-assisted charge transfer to the $\mathrm{CuO_2}$ plane.

\end{abstract}


\maketitle

\section{Introduction}

A noted characteristic of cuprates is anomalies in their phonon spectra in a range below 100 meV. Often with distinct, unusual O-isotope dependence, these have been observed in infrared (IR) \cite{1,2,3}, Raman \cite{4}, neutron scattering \cite{5}, photoemission \cite{6,7,8,9}, and resonant inelastic x-ray spectroscopies \cite{10,11,12}. In such a complex, disordered material, the descriptions of the electron-lattice coupling provided by the unusual behaviors of the peak energies and shapes are insufficient to assist in the development of microscopic models of the superconductivity and other properties of interest. Insofar as the vibrations of the atoms that produce the spectra are defined by the potentials between the atoms, a more direct measure of these anharmonic potentials is found in a real space conjugate to the phonon spectra, namely a snapshot of the atom positions over a large volume of the lattice revealed in the pair distributions. Since these distributions are determined by the underlying pair potentials, not only the presence but also the type and extent of anharmonicity are visualized in the pair distribution functions of the dynamic structure factor, $S(Q,E)$. This quantity is easily intuited in liquids, where the positions of atoms in rapid transit between quasi-stable relative locations are identified in the van Hove function derived from inelastic scattering data \cite{13}. It also occurs in solids in the analogous movements of atoms between locations of similar energies, subject to the constraint that their paths must avoid collisions with the other atoms at their more fixed locations in the solid.

One of the many unusual characteristics of cuprates is the two-site copper-oxygen distributions that are constituents of their dynamic structure. Inelastic neutron scattering measurements identified these shortly after the initial discovery of high temperature superconductivity \cite{14,15}. However, Extended X-ray Absorption Fine Structure (EXAFS) spectroscopy has been the predominant source of these observations because of the relative ease of its measurement of the instantaneous structure factor, $S(q,t=0)$ \cite{16,17}, with a precision that matches the deviations of $S(Q,E)$ from the static structure factor, $S(Q,E=0)$. These two-site distributions indicative of anharmonic, double well potentials were initially observed in the apical $\mathrm{Cu-O_{ap}}$ pairs \cite{18,19,20,21,22,23,24}, then in the planar oxygen ($\mathrm{O_{pl}}$) of the $\mathrm{CuO_2}$ planes in other cuprates \cite{25,26,27,28,29,30,31,32}. This has been followed by reports of similar behavior in superconducting bismuthates \cite{33,34} and even pnictides \cite{35}. Their observed fluctuations at and through the transitions demonstrated their coupling and even their contribution \cite{36} to the superconductivity. More recently, experiments on overdoped superconducting cuprates have revealed the coupling of the highly disordered Cu-Sr pair, and massive changes in the dynamic structure of Cu-O pairs at the superconducting transition of $\mathrm{Sr_2CuO_{3.3}}$ \cite{37}. It is likely that this critical aspect of two-site distributions is a unifying element of high temperature superconductivity and an indication of strong, non-adiabatic, complex electron-phonon/lattice coupling.

\begin{figure}
\includegraphics[width=\linewidth]{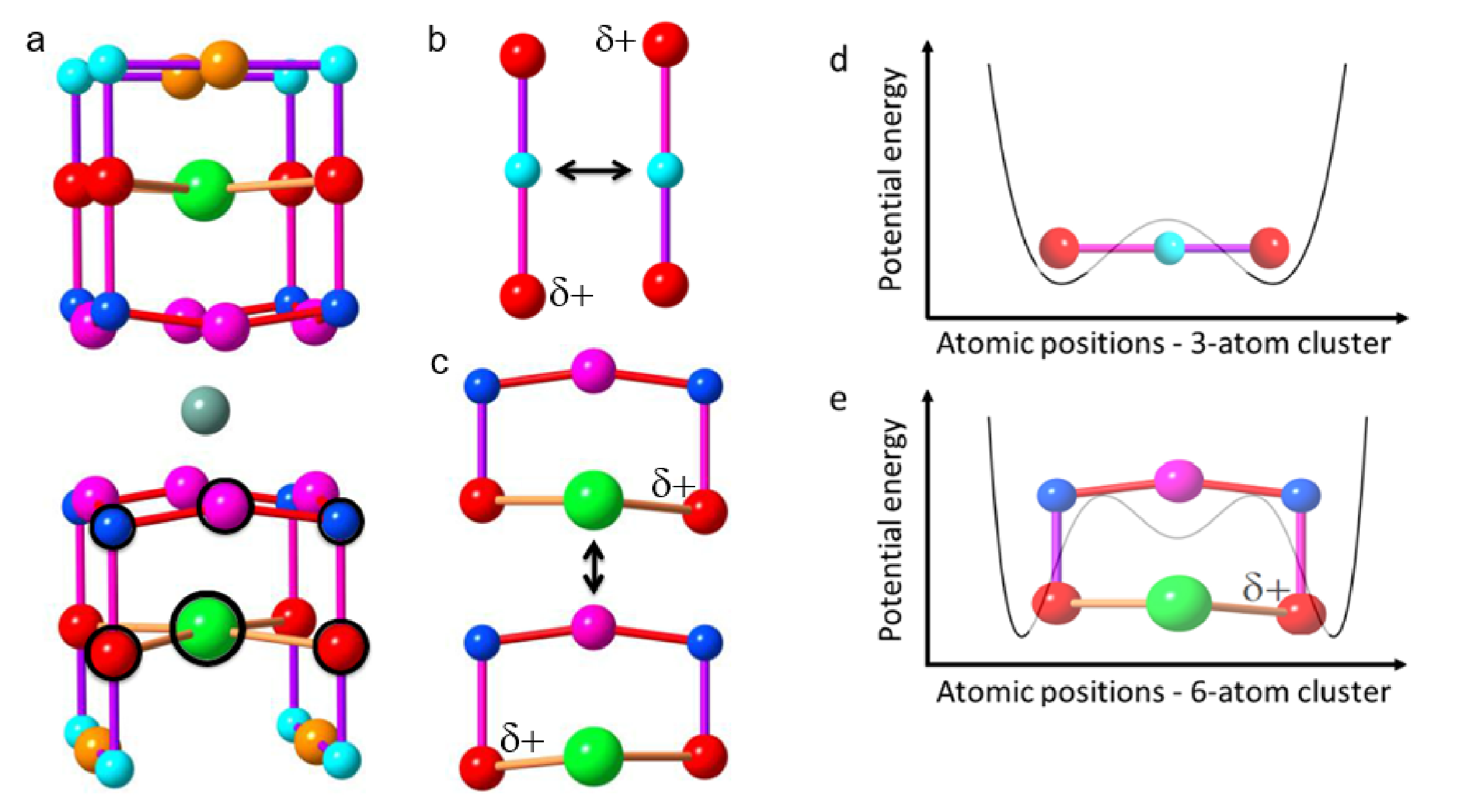}
\caption{Relevant structures. a) The $\mathrm{YBa_2Cu_3O_7}$/$\mathrm{YSr_2Cu_3O_7}$ crystal structure shows the features of multilayered cuprates with higher Tc’s: two, conducting $\mathrm{CuO_2}$ planes (blue Cu(2) and magenta $\mathrm{O_{pl}}$ atoms) bridged by the intervening Y (grey), the charge reservoir consisting in this class of materials of Cu-O chains (turqoise Cu(1) and orange O), and the dielectric layer composed of the $\mathrm{O_{ap}}$ (red) and Sr (green). b) The original three atom IQTP derived from the two-site $\mathrm{Cu(1)-O_{ap}}$ distribution found in $\mathrm{YBa_2Cu_3O_7}$ showing its oscillations between its two configurations denoted by the location of the extra hole ($+\delta$) and expanded Cu-O distance \cite{17}. c) The atoms circled in black in (a) form the six atom cluster used here. Its excess charge and displacement tunnel between the two $\mathrm{Cu(2)-O_{ap}}$ pairs through the $\mathrm{O_{pl}}$ charge-transfer-bridge. d) For the three-atom cluster the potential energy corresponds to a double-well structure \cite{21}. e) For the 6-atom cluster, the potential energy corresponds to a triple-well structure. }
\label{Fig: 1}
\end{figure}

We describe below how the two-site distributions can originate as features of small polarons caused by the doping and mixed-valence character of the materials. Small polarons are defined as a charge inhomogeneity around a central atom and the accompanying local lattice distortion as the neighboring atoms shift their positions to accommodate the different charge. One of their central features is their thermally activated or quantum center of mass tunneling through the lattice. The two-site distribution would occur as a subset of small polarons when the highest occupied states of the system have significant oxygen 2p density. In this case, instead of the hole being confined to the central metal, a fraction of it will reside on one of its neighboring oxygen atoms. This would be expected to result in a second oxygen position at a somewhat longer distance, giving the two-site distribution without perturbing the crystal structure. In the typical transition metal oxide where the metal has equivalent oxygen nearest neighbors, these oxygen ions could exchange the charge and bond length among themselves by temperature-independent quantum tunneling. While this process does shift the site where they reside, in contrast to the small polaron that roams through the lattice, the locations of these modified oxygen atoms are constrained to the nearest neighbor shell of the metal associated with the excess charge. The crucial point is that this process occurs within the parent polaron, constituting internal dynamics in contrast to the lattice dynamics of the parent. When this tunneling frequency is higher compared to the hopping frequency of the parent polaron what we term an Internal Quantum Tunneling Polaron (IQTP) is obtained.

In our earlier work on this IQTP problem we used exact diagonalization calculations for a minimal, three-atom, O-Cu-O cluster with an excess hole (Fig. 1b) \cite{16, 21,22,23,38} to confirm the experimental signature of IQTPs via the difference between the crystallographic, static structure and probes of the dynamic structure that exhibit the two-site distributions. These results validated both the experimental results and the application of such calculations for incorporating local quantum tunneling dynamics. We noted that dynamical contributions were also added in calculations of fluctuating stripes in the $\mathrm{CuO_2}$ planes \cite{39} and in charge flux among the apical cation, apical anion, and the in-plane $\mathrm{CuO_4}$ unit \cite{40}.

However, these reports do not address the full anharmonicity and its connection with the double-well potential in the IQTPs. We do so here via two expansions of the original three-atom calculations. Since the most common moiety is not the linear, O-Cu-O cluster found only in single layer cuprates, but is a Cu surrounded by the $\mathrm{O_{pl}}$ atoms with a single $\mathrm{O_{ap}}$ (Fig. 1a), we incorporate the requisite two $\mathrm{O_{ap}}$ atoms by including two, neighboring $\mathrm{Cu-O_{ap}}$ pairs. Adding the $\mathrm{O_{pl}}$ atom that bridges them incorporates the $\mathrm{CuO_2}$ planes. Finally, attending to the finding of the anharmonicity of the $\mathrm{Sr^{2+}}$ alkaline earth dication \cite{41} and its coupling to the superconductivity \cite{37, 42}, we include this link to the apical oxygen sites that completes the dielectric layer (Fig. 1c). This six-atom cluster derived from the combined structures of $\mathrm{YSr_2Cu_3O_{7+\delta}}$, $\mathrm{YSr_2Cu_{2.75}Mo_{0.25}O_{7.54}}$, and $\mathrm{YBa_2Cu_3O_8}$ now contains most of the functionality of the cuprates and enables our elucidation of the couplings between them. Our calculations show that, under certain conditions, the addition of the planar site, $\mathrm{O_{pl}}$, and a soft, molecular $\mathrm{O_{ap}-Sr-O_{ap}}$ mode causes the double well potential of the $\mathrm{O_{ap}}$ positions in the three-atom cluster (Fig. 1d) to enlarge to a triple well that now involves the $\mathrm{O_{pl}}$ (Fig. 1e), where the depth of the middle well is controlled by the Sr-related anharmonicity.

Our six-atom cluster has been evaluated from four different perspectives. First, we exactly diagonalized the associated quantum many-body Hamiltonian incorporating these additional structural ingredients. We find that the anharmonicity related to the unusual dynamical structure observed in the EXAFS spectrum of Sr/Ba-based cuprates \cite{41,42} is due to its vicinity to a first order synchronization transition of the IQTPs correlated with the two-site $\mathrm{Cu(2)-O_{ap}}$ pair distribution. Second is the realization that an advantageous approach to what is now effectively a system of oscillators is adapting the Kuramoto treatment of networks \cite{43}, which has not previously been attempted in a crystal because of the complex network of couplings through the multiplicity of phonons and because it is an application to a quantum system. In order to provide physical meaning to such a first order transition, we mapped the combined, linearised Heisenberg’s equations of motion for the anharmonically coupled phonons to a mean field Kuramoto equation describing the synchronization of IQTPs within the cluster. The mapping includes both the IQTP anharmonic coupling as well as temperature/dissipation (through thermal disorder) and yields a first order synchronization transition to the anti-phase motions of the two IQTPs. Surprisingly, we find that in the synchronized phase charge is pumped from the apical positions into the conducting $\mathrm{CuO_2}$ plane while, simultaneously, a Sr-related, triatomic molecular vibration develops a finite projection also in the copper oxide plane. This is a novel planar IQTP, associated with $\mathrm{Cu-O_{pl}}$ deformations. Third, we have confirmed the numerical results by performing a multimodal, nonlinear Bogoliubov transformation and demonstrated a polaronic degree of freedom associated with the coupling between excess charge in the plane and a new, anharmonicity-related, lattice vibration. Finally, we have observed that all of our theoretical and numerical results can be summarized in terms of an effective quantum mechanical triple-potential-well model (Fig. 1e). This represents an anharmonic structural adiabatic passage (ASAP) promoting anti-phase IQTP synchronization and internal charge transfer. 

The paper is organized as follows: In Sec. II we describe the here proposed extension of the three-atom into the six-atom cluster and the numerical methodology used to diagonalize its Hamiltonian. In Sec. III the numerical results are presented and discussed within the approach of the Kuramoto model for synchronization of the IQTP’s, we present the non-linear multimodal Bogoliubov transformation and the triple- well interpretation. Finally, Sec. IV is devoted to the conclusions. 

\section{Exact diagonalization} \label{sec:ExactDiagonalization}
We started by performing the exact diagonalization for the extended 6-atom cluster shown in Fig. \ref{Fig: 1}c). As discussed previously, the cluster was carefuly chosen to include two important structural ingredients: (i) a $\mathrm{Cu-O_{pl}-Cu}$ charge transfer bridge, associated with a nearby, planar oxygen ($\mathrm{O_{pl}}$) atom, promoting the transfer of the extra hole and longer $\mathrm{Cu-O_{ap}}$ distance between the two lateral $\mathrm{Cu-O_{ap}}$ IQTPs through the $\mathrm{CuO_2}$ plane; and (ii) a $\mathrm{O_{ap}-Sr-O_{ap}}$ triatomic molecule, associated with a nearby $\mathrm{Sr}$ atom, promoting the anharmonic coupling, referred to as $K$, between the apical oxygens locations of the two lateral $\mathrm{Cu-O_{ap}}$ IQTPs through the non-charge transfer $\mathrm{Sr}$ atom. Notice that while the transfer included in (i) favours charge delocalization, the formation of a tri-atomic molecule, included in (ii), favours the locking of the phases of vibration above a critical anharmonicity $K_c$. As a result of the above rich structure, it is natural to expect that, as anharmonicity is fine tuned accross the cluster, regimes where the excess charge becomes delocalized and vibrations become synchronized are not only to be expected, but, as we are about to show, remarkably interconnected. This process ultimately provides meaning to the anharmonicity-related data observed in EXAFS as discussed previously.

The intricate interplay between lattice and charge degrees of freedom described above can be captured
by the following Hamiltonian $H = H_{el} + H_{ph} + H_{el-ph} + H_{Sr}$ composed of four terms

\begin{align}
H_{el} =\sum_{i}\varepsilon_{i}n_{i}+t\sum_{\left\langle ij\right\rangle\sigma}c_{i\sigma}^{\dagger}c_{j\sigma}+h.c.+ U\sum_{i}n_{i\uparrow}n_{i\downarrow},\label{eq: el}
\end{align}
\begin{align}
H_{ph} = \hbar\omega b_{L}^{\dagger}b_{L}+\hbar\omega b_{R}^{\dagger}b_{R},\label{eq: ph}
\end{align}
\begin{align}
H_{el-ph} =\lambda n_{L}\left(b_{L}^{\dagger}+b_{L}\right)+\lambda n_{R}\left(b_{R}^{\dagger}+b_{R}\right),  \label{eq: elph}
\end{align}
\begin{align}
H_{Sr} = \hbar\Omega_{Sr}\beta^{\dagger}\beta+K\left(\beta^{\dagger}b_{L}b_{R}+\beta b_{L}^{\dagger}b_{R}^{\dagger}\right).\label{eq:sr}
\end{align}
%



\begin{figure*}[!t]
\centering
\includegraphics[width=1\linewidth]{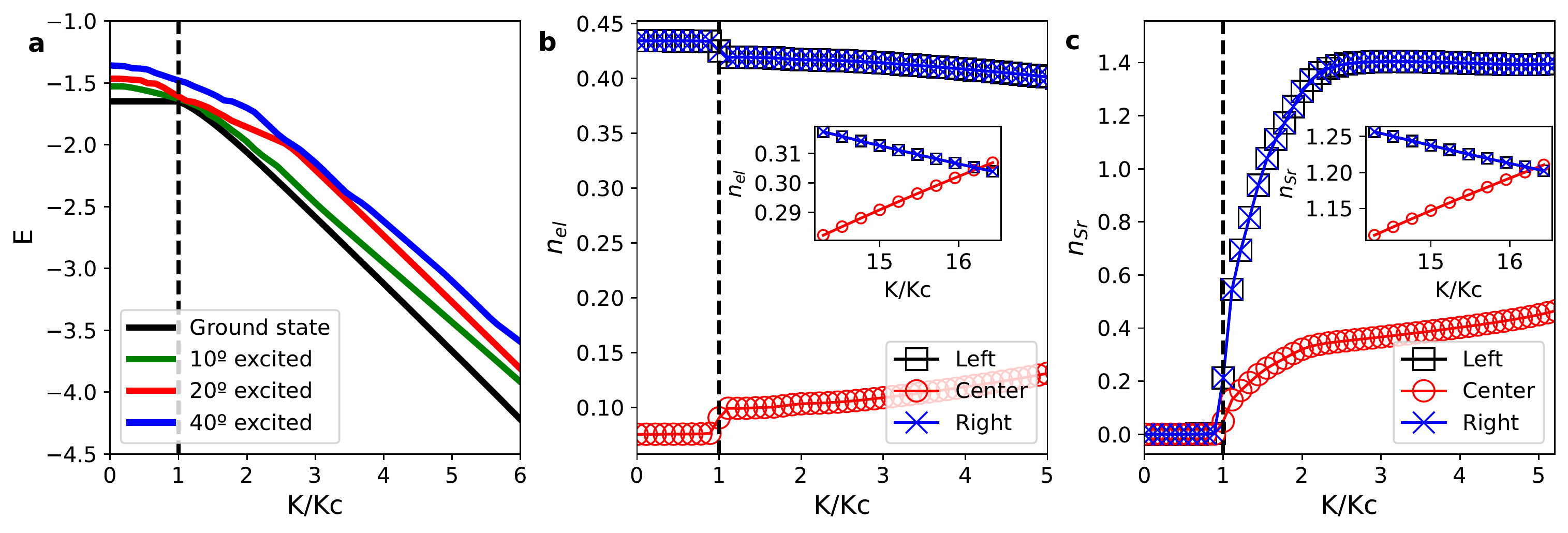}
\caption{{\it a)} The ground state energy of the 6-atom cluster Hamiltonian as a function of reduced anharmonicity, $K/K_c$, for the non-adiabatic, intermediate $\lambda$ coupling regime where lateral polarons are already formed –ground state (black), 10th excited state (green), 20th excited state (red), 40th excited state (blue). {\it b)} The electronic occupations for left/right $\mathrm{O_{ap}}$ (black/blue squares/crosses) as well as for the central $\mathrm{O_{pl}}$ (red circles), as a function of reduced anharmonicity, $K/K_c$ of the ground state. {\it c)} The phonon occupations for left/right sites (black/blue square/cross) as well as for the central oxygen site (red circles), as a function of reduced $K/K_c$ anharmonicity, of the ground state. {\it Insets:} as we increase anharmonicity, the central electronic (b) and phonon (c) occupations surpasses the lateral ones.}
\label{Fig: 2}
\end{figure*}

In the electronic part Eq. \eqref{eq: el}, $c^\dagger_{i,\sigma}$, $c_{i,\sigma}$ are the usual creation and annihilation operators for holes with spin projection $\sigma$, with $i = 1, …, 5$ representing the $\mathrm{O_{ap}-Cu-O_{pl}-Cu-O_{ap}}$ charger transfer sites, respectively, $n_{i} = \sum_{\sigma} c^\dagger_{i,\sigma}c_{i,\sigma}$ is the number occupation operator for holes at sites $i$, where $n_1 = n_L$ and $n_5 = n_R$, with on-site energies $\varepsilon_i$, $t$ is the spin-preserving, nearest-neighbour hopping amplitude, and $U$ the on-site Coulomb repulsion to prevent double occupancy. The site energies were chosen so that out of a total of three holes, as considered in this work, two of them will always be favoured at the two $\mathrm{Cu}$ atomic positions, while only a single, remaining excess hole minimizes the total energy by moving among the oxygen atoms while avoiding the two copper atomic positions. The phonon part Eq. \eqref{eq: ph} consists of the two harmonic infrared oscillators, of identical normal frequencies $\omega$ whose creation and annihilation operators $b^\dagger_L$, $b_L$, and $b^\dagger_R$, $b_R$ represent the two possible lateral $\mathrm{Cu-O_{ap}}$ positions, to the left (L) and to the right (R), in the cluster. The electron-phonon interaction term Eq. \eqref{eq: elph} describes the local coupling between the hole degrees of freedom to the lattice displacements, controlled by the coupling constant $\lambda$. This type of local electron-phonon coupling (Holstein model) is chosen due to its simplicity compared to non-local couplings (Su-Schrieffer-Heeger model) and because the same physical picture underlies the formation of a single polaron in the two cases \cite{44}. The novelty here is Eq. \eqref{eq:sr}, which was designed to provide the cluster with a nontrivial, $\mathrm{O_{ap}-Sr-O_{ap}}$ triatomic molecule structure, motivated by the coupling of the anharmonicity of the neighboring $\mathrm{Sr}$  to the behavior of $\mathrm{O_{ap}}$ atoms and the superconductivity. To describe the normal modes of vibration of the triatomic molecule we have introduced creation (annihilation) operators $\beta^\dagger$ ($\beta$) for a moderately stiff harmonic molecular phonon of normal frequency, $\Omega_{Sr} = 2 \omega$, as required by energy conservation. As we can see from Eq. \eqref{eq:sr} the new interaction term between the molecular phonons and the rest of the cluster is indeed anharmonic, in the form of a three-phonon coupling, and controlled by an anharmonic coupling, $K$. This interaction term can be interpreted as a molecule formation term, and $K$ as a chemical potential that controls processes in which left and right phonons are destroyed to form a molecular mode, $\beta^\dagger b_L b_R$ , as well as a destruction of a molecular vibration to produce left and right independent oscillations, $\beta b^\dagger_L b^\dagger_R$. As such, one would naturally expect that a molecule formation interaction would work towards promoting the locking of phases for the $\mathrm{O_{ap}}$ vibrations, or equivalently, the synchronization of Kuramoto oscillators. For this reason the anharmonic coupling, $K$, will be referred to as the Kuramoto coupling.

The exact diagonalization of the full Hamiltonian, for the ground and excited states, was perfomed using a basis of wavefunctions given by

\begin{equation}
    |\Psi\rangle=\sum_{i,\gamma,\beta,\delta}\alpha_{i\gamma\beta\delta}
    \left|n_i\right>\left|n_\gamma\right>\left|n_\beta\right>\left|n_\delta\right>,
\end{equation}
corresponding to a $n_{el}\times n_L \times n_R \times n_{Sr}$ dimensional Hilbert space. For the electronic states $|n_i\rangle$ we first considered all possible spin-up and spin-down configurations at the five charge transfer sites and a total of three holes added to the cluster. However, as discussed above, since only a single (excess) hole is found over the five charge transfer sites while the other two are favoured at the copper sites, our electronic subspace was limited to $n_{el} = 28$ states. The bosonic $|n_\gamma \rangle | n_\beta \rangle |n_\delta\rangle$ states represents the phonon degrees of freedom for the left and right apical positions and for the Sr-related triatomic molecule, respectively. These states were written in a bosonic occupation number representation for a total of $n_L = n_R = n_{Sr} = 5$, justified by continuosly enlarging the phonon Hilbert subspace, through a systematic increase of the number of phonon modes until convergence.

For the numerical parameters in the Hamiltonian \eqref{eq: el}-\eqref{eq:sr} we used \cite{23} $\varepsilon_{1,3,5}=-\varepsilon_{2,4}=0.5\; eV$, favouring holes at the Cu atoms, $t=1.0  \;eV$, and $U=7.0 \; eV$. For the electron-phonon coupling, we set $\lambda = 0.3\; eV$, inside the non-adiabatic regime for polaron formation \cite{45} and we have chosen the stiff, triatomic molecule frequency, $\Omega_{Sr}=0.12 \; eV$, to be precisely twice the value of $\omega=0.06 \;eV$, as required by energy conservation.

\section{Numerical Results and discussion}\label{Sec:numerical}

\begin{figure*}[!t]
\centering
\includegraphics[width=1\linewidth]{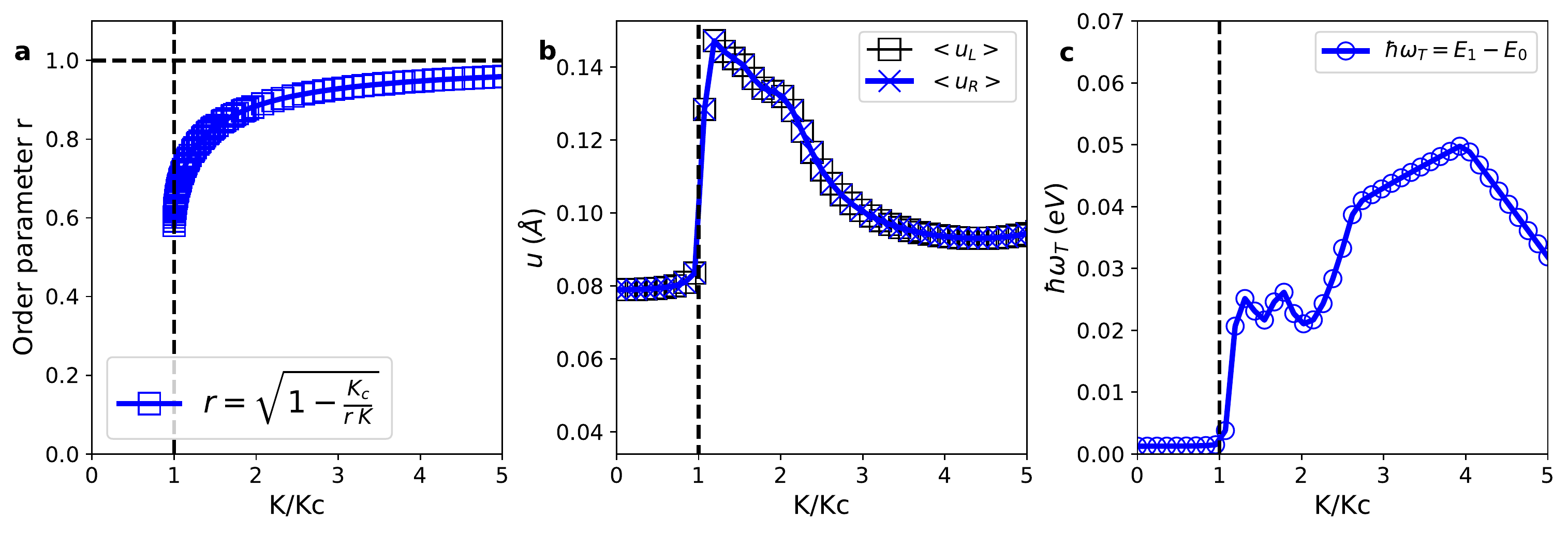}
\caption{{\it a)}: The solution to Kuramoto’s order parameter equation (\ref{eq: 7}) as a function of the reduced coupling in the ground state, showing a first order synchronization phase transition. {\it b)}: Displacement $<u> = \sqrt{\hbar/2m\omega}(b + b^\dagger)$ of the $\mathrm{O_{ap}}$ equilibrium position for both the left and right positions in the cluster. {\it c)}: The polaronic tunneling frequency.}
\label{Fig: 3}
\end{figure*}

In Fig. \ref{Fig: 2} we summarize the results obtained from our numerical, exact diagonalization studies. Figure \ref{Fig: 2}a) shows the ground and three representative excited states energies as a function of the reduced coupling, $K/K_c$. For the ground state (black line $-$ first from bottom up) one observes a clear kink at $K/K_c= 1$ while for all higher excited states (green, red, and blue lines $-$ second, third and fourth from bottom up) there is no clear kink, although a decrease in energy is evident even for $K<K_c$. We can see that for the higher excited states, the definition of a critical coupling is not as clearer as for the ground state because of level avoiding behavior. Since higher excited states will play a significant role only when $T \neq 0$, we restrict to the behavior of the ground state across the synchronization transition. Figure \ref{Fig: 2}b) shows the electronic occupations at the left (black squares), central (red circles), and right (blue crosses) sites, also as a function of the reduced coupling, $K/K_c$, for the ground state. At the critical coupling, $K_c$, we have found a delocalization transition of the excess hole, as anticipated. In fact, while prior to the transition the hole wave function had weight only at either the left or right apical oxygen atoms, indicating formation of localized polarons, above $K_c$ the electronic occupation for the central (planar) oxygen atom begins to increase monotonically while both the left and right occupations decrease. Indeed, when $K\gg K_c$, the central occupation is greater than the apical ones. The abrupt discontinuity demonstrates that the phase transition is first order and this behavior indicates the formation of a {\it split-polaron.} Finally, Fig. \ref{Fig: 2}c) shows the occupations of $\mathrm{Sr}$-related, molecular phonons at the left (black squares), right (blue crosses), and central (red circles) positions within the cluster. When $K$ crosses $K_c$, all phonon modes related to the triatomic molecule become active, which is expected from the apical oxygens' phase locking driven by the synchronization. But remarkably, as shown in the inset, the increase of anharmonicity drives the central projection of molecular phonons to be greater than the lateral ones.

The numerical results reveal three important features associated with the Hamiltonian (\ref{eq: el})-(\ref{eq:sr}) as anharmonicity, encoded in the Kuramoto parameter $K$, is varied, Fig. \ref{Fig: 3}: (i) as shown in section III-A, it produces a first order synchronization phase transition associated with the anharmonicity-related, triatomic molecular locking or synchronization of the phases of vibration corresponding to the two $\mathrm{O_{ap}}$ locations or $\mathrm{Cu-O_{ap}}$ distances when the Kuramoto coupling $K$ is increased, evidenced by the evolution of the Kuramoto order parameter, $r$, with anharmonicity (Fig. \ref{Fig: 3}a); (ii) a non-linear behavior for the displacement $\langle u \rangle$ related to the apical oxygens, which is quantitatively in agreement with the experimental deviations found in the two-site distributions associated with the $\mathrm{Cu-O_{ap}}$ pairs \cite{46} (Fig. \ref{Fig: 3}b). Before the transition, there is a stable $|\langle u \rangle| \approx 0.08 \; \AA$ position, since $\lambda$ is strong enough to form localized, frozen polarons, and although a dynamical shift occurs after the synchronization (a signature of the non-linear behavior), there's still agreement with EXAFS measurements of the distance in the $\mathrm{Cu-O_{ap}}$ pair \cite{45}; (iii) it generates a polaronic tunneling frequency, $\hbar \omega_T = E_1 - E_0$ that is the difference in energy from the first to the ground state \cite{22}, that evolves nontrivially with $K$ (Fig. \ref{Fig: 3}c). There is a jump at the critical coupling, indicating the value of the anharmonicity beyond which the initially frozen polarons become an IQTP, allowing quantum tunnel internally between the two lateral positions in the cluster. In order to address each one of these issues, we proceed by applying three different techniques.

\subsection{The first order synchronization phase transition}

Collective behavior and its spontaneous emergence in networks of coupled oscillators is a common characteristic in many systems across science and perhaps the simplest model that describes the synchronization phenomena is the Kuramoto model \cite{47}. Its application ranges from classical systems in physics \cite{48} and biology \cite{49} to quantum physical systems \cite{50,51,52}. The Kuramoto model is based on two properties: the node oscillators are coupled via a superfluid density $K_{ij}$ and there are white, quenched, $\delta$ or thermal, $k_B T$, noises that are present due to the environment the networks are embedded in. Synchronization can be achieved whenever the couplings predominate over the noises. The synchronization process is described by a complex order parameter, $re^{i\psi}$, where the real part $r$ sets the character of the synchronization: $r=0$ for the unsynchronized phase, $0<r<1$ for partial and $r=1$ for full synchronization (see Fig. \ref{Fig: 3}a)). The phase $\psi$ in the imaginary part of the order parameter sets the overall phase that the oscillators achieve when synchronized.

As demonstrated in Appendix \ref{sec:appendix A}, the set of coupled equations of motion for the vibrational degrees of freedom obtained from the Hamiltonian (\ref{eq: el})-(\ref{eq:sr}) can be conveniently rewritten, within a mean field approximation \cite{52, 53}, as a single, first–order differential equation for the phases of the lateral oscillators, $\theta_i$, in which temperature effects can be added by introducing a white thermal noise, $\zeta_i(t)$, thus providing, in terms of the Kuramoto’s complex order parameter $re^{i\psi}$

%
\begin{eqnarray}
\dot{\theta}_i=\omega_{i}+K r^{2} \sin \left[\theta_{i}(t)+\psi\right]+\xi_{i}(t),
\end{eqnarray}
such that $\left\langle \zeta_i(t) \right\rangle = 0$ and $\left\langle \zeta_i(t)\zeta_j(t^\prime)\right\rangle = 2\gamma k_B T\delta_{ij}\delta(t-t^\prime)$, where $\gamma$ is a damping constant. The solution to such first order equations yields
\begin{eqnarray}
r = \sqrt{1 - \frac{K_c(\delta, T)}{rK} },
\label{eq: 7}
\end{eqnarray}
unveiling the connection between the Kuramoto order parameter and the anharmonicity introduced in the cluster. It also shows the first order nature of the synchronization transition that occurs at $K_c(\delta, T)$, a critical coupling  (see Eq. \ref{eq:ap_10}) that is determined both by the temperature $T$ as well as a quenched spread $\delta$ that relates to disorder. In summary, the process of increasing anharmonicity induces the localized polaronic phase $(K<K_c)$ to evolve into a phase of synchronized internal tunneling polarons, since $\hbar \omega_T \neq 0$ for $K> K_c$ (Fig. \ref{Fig: 3}c)), changing dynamically the otherwise stable position of the $\mathrm{Cu-O_{ap}}$ pairs (Fig. \ref{Fig: 3} b)).

\subsection{The formation of a planar IQTP}

The exact diagonalization results show that not only the total Sr-phonon occupation jumps at $K_c$ but also, most interestingly, an unexpected Sr-phonon central occupation, $n_\beta (C)$, appears above the first order anti-phase synchronization transition. This observation motivated us to {\it break down} the $\mathrm{O_{ap}-Sr-O_{ap}}$ triatomic molecule term of the original Hamiltonian \eqref{eq:sr} as
\begin{equation}
H_{Sr}=\sum_{r}\hbar\Omega_{Sr}\beta_{r}^{\dagger}\beta_{r}+K\sum_r \left(\beta_{r}^{\dagger}b_{L}b_{R}+\beta_{r}b_{L}^{\dagger}b_{R}^{\dagger}\right),
\end{equation}
where $\beta_r,\beta_r^\dagger$ are now to be understood as the projection of the triatomic molecular vibration at the relevant sites of the cluster, namely $r=L,C,R$.

In order to diagonalize the Hamiltonian and connect the numerical results showing the appeareance of an excess charge, together with phonon projections in the central, planar site of the cluster, which leads to the formation of a planar IQTP, we introduce a non-linear, multimodal Bogoliubov transformation for the lateral phonons as

\begin{eqnarray}
b_{L} & \;=\; & \sum_{r}\left(u_{L,r}^{*}\beta_{r}B_{r}+v_{L,r}\beta_{r}^{\dagger}B_{r}^{\dagger}\right),\; \;\; \; \; r = R, C\\
b_{R} & = & \sum_{r}\left(u_{R,r}\beta_{r}^{\dagger}B_{r}^{\dagger}+v_{R,r}^{*}\beta_{r}B_{r}\right), \;\;\;\;\; r = L,C
\end{eqnarray}
together with the complex conjugate $b^\dagger_L$ and $b^\dagger_R$. We emphasize that by transforming the Hamiltonian from the lateral phonon operators to the new bosonic $B_r$ modes, we are searching for the limit where the charge and phonon projections at the central site $(r=C)$ are interacting. This is achieved by taking the limit of strong anharmonicity $K\gg K_c$, as suggested by the numerical results in the insets of Fig. \ref{Fig: 2}b) and Fig. \ref{Fig: 2}c). Therefore, after applying the Bogoliubov transformation procedure in this limit (see Appendix \ref{sec:appendix B}), the total diagonal phonon Hamiltonian in terms of new Bogoliubov phonons can be written as 
\begin{eqnarray}
H_{d}^{\prime}= \hbar F(K) B_{C}^{\dagger} B_{C},
\label{eq:9}
\end{eqnarray}
and anharmonicity-dependent, novel, central phonon modes are present, whose natural frequencies $\hbar F(K)$ are given by
\begin{eqnarray}
\hbar F(K) & = & \frac{\hbar\omega}{A\gamma} \left[\omega(1+2n_\beta(C)) - \gamma\right]\nonumber\\
& + & \frac{K\sqrt{n_\beta(C)}}{A\gamma} \sqrt{\omega^2 - \gamma^2}(1+2n_\beta(C)).
\end{eqnarray}
where $n_\beta (C)$ is the occupation of Sr-related triatomic molecular phonons projected on the central site (red circles in Fig. \ref{Fig: 2}c)), $\hbar\gamma = \sqrt{(\hbar\omega)^2 - K^2n_\beta(C)}$ and $A=1+n_\beta (C)+n_B (C)$. We have found that, before synchronization, $K<K_c$, $\hbar F(K) = 0$ because $n_\beta(C) = 0$, thus these new phonon modes are only present in the limit of strong anharmonicity $K\gg K_c$. The multimodal, nonlinear Bogoliubov transformation needs also to be applied to the original electron-phonon coupling, which together with equation (\ref{eq:9}), gives rise to the new transformed Hamiltonian, that can be written as
\begin{eqnarray}
H_{S r}^{\prime}& = & \hbar F(K) B_{C}^{\dagger} B_{C} + \lambda^{\prime} n_{ap}\left(B_{C}+B_{C}^{\dagger}\right),
\label{eq: 13}
\end{eqnarray}
where $n_{ap}$ is to be understood as the electronic occupation at the apical positions and the new excess-hole-central-phonon coupling is given by
\begin{eqnarray}
\lambda^\prime = \frac{2\lambda \sqrt{n_\beta (C)}}{\sqrt{2A\gamma}}\left(\sqrt{\omega + \gamma} + \sqrt{\omega - \gamma}\right).
\end{eqnarray}

This is a striking result and demonstrates that after the synchronization of the left and right $\mathrm{O_{ap}}$ vibrations, the transition to a synchronized IQTP’s phase, with the pumping of charge and phonon projection to the $\mathrm{CuO_2}$ plane, promotes the formation of an IQTP in the central, planar site, as suggested by the numerical results. Therefore, our analysis has demonstrated that a planar IQTP arises from the synchronization transition,  providing theoretical support to the observation of IQTPs in the planar oxygen ($\mathrm{O_{pl}}$) of the $\mathrm{CuO_2}$ planes in some compounds \cite{25,26,27,28,29,30,31,32}, and even in other HTSC materials \cite{33,34,35}.

The electron-phonon part, $H_{e l-p h}^{\prime}$, of the transformed Hamiltonian, second term in Eq. (\ref{eq: 13}), can also be written in a way to show the explicit interaction between the central electronic occupation and the new phonon modes present at the same site. Since only one excess hole is added to the system, the constraint $n_L + n_R + n_C = 1$, or even $2n_{ap} + n_C = 1$, is valid, thus $n_{ap} = (1-n_C)/2$. Therefore, the Bogoliubov analysis supports the interpretation of the triple-well structure arising after the IQTPs synchronize, which we explore in the next section. 

\begin{figure*}
\centering
\includegraphics[width=1\linewidth]{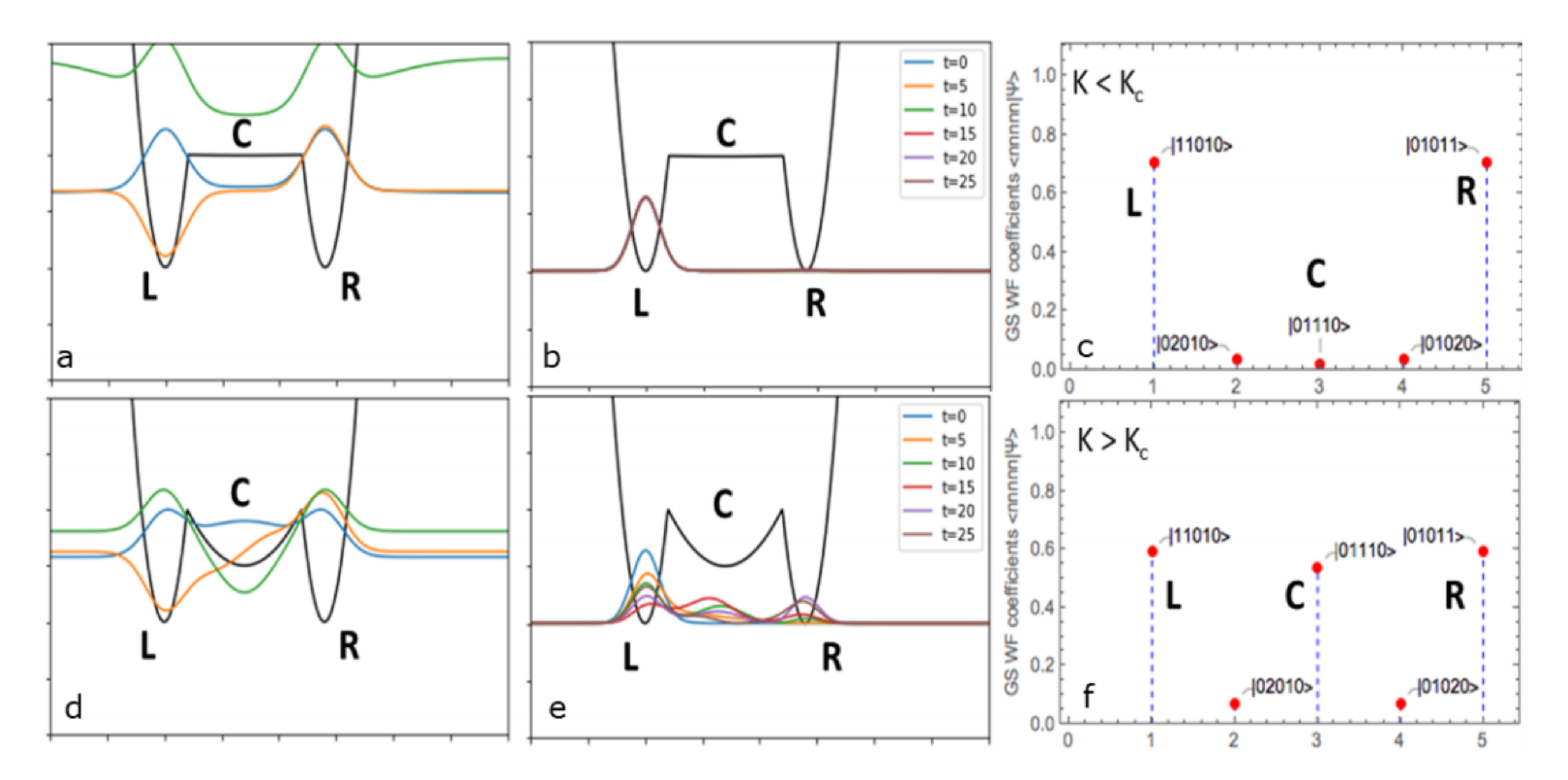}
\caption{a) For $K<K_c$, the potential (black) depth, the symmetric (blue) and anti-symmetric (orange) components of the lateral ground state wave function and a central excited state (green). b) The absence of tunneling of the polarons in this scenario is demonstrated by its fixed position over time. c) Ground state wave function coefficients for the unsynchronized, localized polarons phase. d) for $K>K_c$, the triple-well potential (black), the symmetric (blue) and anti-symmetric (orange) solutions of the wave function and the excited state (green). e) Due to the ASAP, the IQTP is now observed to tunnel from left to right through the center as time evolves. f) Ground state wave function coefficients for the synchronized IQTPs phase. The states in c) and f) are represented in the electronic occupation representation for the three holes in the five charge transfer sites.}
\label{Fig: 4}
\end{figure*}

\subsection{The triple-well}

The triple-well structure can now be elucidated. Before synchronization, $K<K_{c}$, we have $n_{\beta}(C)=0$ and $n_{el}(C) = 0$, thus the system contains two decoupled harmonic vibrations localized at each possible $L, R$ positions, composing a double-well structure (see Fig. \ref{Fig: 4}a)). The potential depth and width are large and the symmetric and anti-symmetric components of the lateral ground state wave function are degenerate, while a central excited state is too energetic to support tunneling. In this case the single excess hole in the cluster is found either at the (L) or (R) apical oxygen atoms. Once the system is prepared at the (L) position, for example, it remains at that position even after time evolves, as shown explicitly in Fig. \ref{Fig: 4}b). This results from the absence of a tunneling frequency for $K<K_{c}$ as shown in Fig. \ref{Fig: 3}c). The ground state wave function coefficients given in Fig. \ref{Fig: 4}c) shows the  polaron localization as finite occupation appears only at the lateral positions, and zero electronic occupation at the central position, for $K<K_{c}.$

After synchronization, $K>K_{c}$, however, we have $n_{\beta}(C) \neq 0$ and $n_{el}(C) \neq 0$. The central site of the system is now active, and when $K\gg K_c$ a new harmonic central phonon mode is present, composing a three-well structure as shown in Fig. \ref{Fig: 4}d). The symmetric and anti-symmetric solutions of the wave function become non-degenerate and the central well on the planar oxygen site in the synchronized phase substantially lowers the energy of the excited state, forming an anharmonic structural adiabatic passage that promotes tunneling between left and right IQTPs. This results in the nonzero tunneling frequency of the synchronized phase. In this case the single excess hole in the cluster can be found at all (L), (R) and (C) oxygen atoms. If the system is prepared at the (L) position the existence of a finite tunneling frequency shows that the wave function evolves with time and the excess hole becomes delocalized. This is depicted in Fig. \ref{Fig: 4}e). The ground state wave function coefficients given in Fig. \ref{Fig: 4}f) shows finite but decreasing occupations at the lateral positions, and nonzero, increasing electronic occupation at the central position, for $K>K_{c}$. Small but finite occupation of the excess hole also occurs on the two $\mathrm{Cu}$ sites, demonstrating delocalization throughout the entire cluster.

\section{Conclusion}

Our exact diagonalization calculations of the original 3-atom O-Cu-O cluster produced significant but limited results on experimental signatures of IQTPs \cite{16,17,21}. In this work, we have expanded the cluster to 6 atoms to incorporate the essential functionalities of the layered cuprate structures: a pair of $\mathrm{Cu-O_{ap}}$ IQTPs, their $\mathrm{O_{pl}}$ bridge that adds the $\mathrm{CuO_2}$ conducting planes, and their divalent alkaline earth cation (Sr) that, with the $\mathrm{O_{ap}}$ atoms, completes the atoms of the dielectric layer. This extension allows a much more detailed investigation of the local interplay between the lattice and electronic degrees of freedom in the cuprates and related compounds. A novel and crucial element in these calculations is the anharmonic, three-phonon coupling in the cluster Hamiltonian whose character and coupling to the superconductivity has been identified by EXAFS measurements on several compounds \cite{37,42,46}. The most notable result reported in this work, that should generalize beyond cuprates to related systems, is the transition to a synchronized phase whose broken symmetry is found in the dynamic structure. This occurs at a critical anharmonic coupling, $K_c$, the Kuramoto coupling between the two IQTPs. Pertinent to the electron-lattice coupling and possible associated superconductivity mechanism in this spontaneous heterostructure, is that in the synchronized phase a fraction of the excess charge that was originally localized as small polarons frozen on the $\mathrm{O_{ap}}$ is transferred to the $\mathrm{O_{pl}}$ of the $\mathrm{CuO_2}$ planes that are the locations of the superconductivity in the oxide-cuprates. A second relevant behavior of the synchronized phase is the projection from the triatomic $\mathrm{O_{ap}-Sr-O_{ap}}$ molecule of phonon modes throughout the cluster. These are controlled by the new phonon $(\beta)$ in the anharmonic term of the Hamiltonian. The displacements of the $\mathrm{O_{ap}}$ atoms from their static positions to form the two-site distribution, $\langle u \rangle$, quantitatively reproduce the $0.12-0.2 \AA$ separations found by EXAFS \cite{46}. This agreement with experiments supports our calculations.

Furthermore, here the Kuramoto model was applied by mapping the Heisenberg equations of motion for each oscillator in the $\mathrm{O_{ap}}$ position to a system of coupled oscillators. By doing so we could describe the transition at $K_c$ as an anti-phase synchronization transition for the phases of each lateral oscillator. For $K>K_c$, the otherwise localized polarons, or independent vibrations, became dynamically active in the form of synchronized IQTPs due to the presence of the stiff molecular vibration brought by the Sr atom in the cluster that connects the two $\mathrm{Cu-O_{ap}}$ pairs. Moreover, a non-linear Bogoliubov transformation was applied to the Hamiltonian in a way that, in the limit of strong anharmonicity, $K \gg K_c$, new phonon modes are present in the planar site. In combination with excess charge pumped into the $\mathrm{O_{pl}}$ of the $\mathrm{CuO_2}$ plane in the synchronized IQTP phase, a new electron-phonon, thus a polaronic, degree of freedom occurs in the plane that can be interpreted as a planar IQTP \cite{25,26,27,28,29,30,31,32}. Together, these theoretical findings points to the triple-well potential interpretation, where the appearance of a finite tunneling frequency in the synchronized phase originates in the pumping of charge from $\mathrm{O_{ap}}$ onto the $\mathrm{O_{pl}}$ site. This lowers the energy of this site and enables the excess charge to quantum tunnel across the cluster, resulting in the internal quantum tunneling polaron dynamics of the system.

Despite the limitations inherent to the small 6-atom cluster, we emphasize that the additional components of this larger cluster and especially the anharmonicity introduced by the Sr atom \cite{41,42} are crucial to understanding the dynamics of lattice and charge degrees of freedom in this system. This points towards the need to understand how anharmonic phonons can influence important characteristics of different systems \cite{54,55,56}, now including cuprates. In this work, we have added a new interpretation of the dynamics related to the collective motion of atoms in the lattice based on the Kuramoto model for synchronization. These results motivate further investigation of the connection between synchronized IQTPs described as Kuramoto oscillators and anharmonic phonons and its role in the charge-lattice dynamics of cuprates, including superconducting properties and the role of Coulomb repulsion, as for example when considering two holes added in an IQTP or the patterning of finite densities of coupled IQTPs \cite{57}. We note that EXAFS measurements of cuprates in which Co, Mn, and Ni were substituted for a fraction of the Cu in the $\mathrm{CuO_2}$ planes demonstrated that the Cu-O IQTPs are not only coupled to but in fact play a direct role in HTSC \cite{36}. Recent reports on highly overdoped superconducting cuprates prepared by high pressure oxygen (HPO) methods have provided exceptions to many of what have been considered common unifying behaviors of cuprate superconductivity, namely: in $\mathrm{Sr_2CuO_{3.3}}$, copper oxide planes as $\mathrm{CuO_{1.5}}$ instead of $\mathrm{CuO_2}$ \cite{37}; continued retention of or increases in $T_c$ throughout their $50-115 \; K$ transition temperatures through excess Cu charge values even beyond 0.6, which far exceed the 0.27 limit of the dome in the conventional phase diagram \cite{46}; a reversal of the correlation between longer $\mathrm{Cu-O_{ap}}$ distances and higher $T_c$, oblate Cu geometry and inversion of the Cu $3d_{z^2-r^2}$ and $3d_{x^2-y^2}$ energies in $\mathrm{Ba_2CuO_{3.2}}$ \cite{58}, and a structural transformation concomitant with the superconducting transition \cite{59}. Insofar as a universal factor is the presence of IQTPs coupled to the superconductivity, our findings on IQTPs signatures in the dynamic structure may play a key role in developing the still unresolved HTSC pairing and condensation mechanism.

\section*{Acknowledgements}

The authors acknowledge the financial support from the Slovenian Research Agency (core funding No. P1-0040). Work at Washington State University is partially supported by the US National Science Foundation Division of Materials Research Early Concept Grants for Exploratory Research Grant 1928874. Use of the Stanford Synchrotron Radiation Lightsource, SLAC National Accelerator Laboratory, is supported by the US Department of Energy, Office of Science, Office of Basic Energy Sciences Contract DE-AC02-76SF00515. VV and MBSN acknowledge the financial support of CAPES and FAPERJ.

\appendix

\section{Kuramoto's synchronization transition} \label{sec:appendix A}
We start by providing meaning to the first order transition. To this end, we shall make use of a mean field approximation to treat the anharmonic three-phonon term in Eq. \eqref{eq:sr}. For convenience, first we rescale the anharmonic coupling $K\rightarrow K/N$, where $N$ represents the number of different oscillators (phonons) at each of the two kinds of oscillator communities: left (L) and right (R) oscillators. Second, since all oscillators are bosons we introduce a mean field approximation in terms of which the triatomic molecular phonon population can be written as $\langle \beta^\dagger \beta\rangle =|R|^2=\langle \beta^\dagger \rangle\langle \beta \rangle=R^* R\neq 0$, where $R$ plays the role of a complex order parameter. Now, the Heisenberg’s equations of motion derived from Hamiltonian \eqref{eq: el}-\eqref{eq:sr} for $i=L,R$ phonons read 
\begin{eqnarray}
i \frac{d b_{i}}{d t} &= &\omega_{i}\left(b_{i}-\frac{\lambda n_{i}}{\hbar \omega}\right)+\frac{K R}{N \hbar} b_{j}^{\dagger}, \nonumber\\
-i \frac{d b_{i}^{\dagger}}{d t}&=&\omega_{i}\left(b_{i}^{\dagger}-\frac{\lambda n_{i}}{\hbar \omega}\right)+\frac{K R^{*}}{N \hbar} b_{j}. \label{eq:ap_1}
\end{eqnarray}
So far, this is a set of coupled operatorial differential equations. Let us take the quantum mechanical expectation value at both sides of the above equations, and recall that phonon coherent states are eigenstates of the annihilation operator $b_{i,j} |z_{i,j}\rangle = z_{i,j} |z_{i,j} \rangle$. We then drop all constant terms that simply provide an overall shift of the equilibrium position and write the complex numbers $z_{i,j}=z_0 e^{-i\theta_{i,j}(t) }$, where we have made the assumption that the amplitudes of the phonons L and R are equal, $z_0$, and only their phases change. Finally we introduce Kuramoto’s complex order parameter, $R=re^{i\psi}$, where $r$ is the real part of $R$ and $\psi$ is an arbitrary phase. Furthermore, we choose $R=-ir$, as purely imaginary, so that after combining the two equations in \eqref{eq:ap_1} we arrive at 
\begin{eqnarray}
\frac{d \theta_{i}(t)}{d t}=\omega_{i}+\sum_{j=L, R} \frac{K r}{N \hbar} \sin \left[\theta_{i}(t)+\theta_{j}(t)\right].\label{eq:ap_2}
\end{eqnarray}
We recognize the dynamical problem in \eqref{eq:ap_2} as a Kuramoto’s differential equation for anti-phase synchronization $\theta_i\rightarrow -\theta_j$. The mean field version of the above equation is obtained as usual, by introducing the order parameter \cite{43} $re^{i\psi}=1/N \sum_{j=1}^N e^{i\theta_j} $. Here the real part $r$ plays the role of the coherence amplitude for a population of $N$ phase oscillators and $\psi$ indicates the coherence phase. Kuramoto’s mean field equation can then finally be written as
\begin{eqnarray}
\dot{\theta_{i}}(t)=\omega_{i}+K r^{2} \sin \left[\theta_{i}(t)+\psi\right]\label{eq:ap_3},
\end{eqnarray}
where from now on we set $\hbar=1$. According to equation \eqref{eq:ap_3}, phase locking occurs for $\theta_i\rightarrow -\psi$, and as such the anti-phase synchronization favoured by equation \eqref{eq:ap_2} implies $\theta_j\rightarrow\psi$. Now, since we have chosen before $\psi=-\pi/2$ (such that $R=re^{i\psi}=-ir$) we end up with $\theta_i+\theta_j=0$ and $\theta_i-\theta_j=-\pi$, which reflects the anti-phase synchronization transition. 

Once we established the anti-phase character of the synchronization transition, we now show its first order nature. For that, the mean field analysis initiated above for the Kuramoto model needs to be supplemented by a self consistency equation \cite{43} such as
\begin{eqnarray}
1=K r \int_{-\frac{\pi}{2}}^{\frac{\pi}{2}} \cos ^{2} \theta g\left(K r^{2} \sin \theta\right) d \theta \label{eq:ap_4},
\end{eqnarray}
where $g(\omega)$ is a symmetric Lorentzian distribution of frequencies with quenched spread $\delta$ given by
\begin{eqnarray}
g(\omega)=\frac{1}{\pi} \frac{\delta}{\omega^{2}+\delta^{2}}\label{eq:ap_5}.
\end{eqnarray}

Solving the self consistent equation, one finds for the order parameter
\begin{eqnarray}
r=\sqrt{1-\frac{K_{c}(\delta)}{K r}} \label{eq:ap_6},
\end{eqnarray}
where $K_c (\delta)=2\delta$ is the critical coupling in terms of a quenched spread $\delta$. 

Finally, since we would like also to incorporate thermal effects to the synchronization transition, we recall that temperature effects can be added to the problem by introducing a thermal noise 
\begin{eqnarray}
\dot{\theta}_i(t)=\omega_{i}+K r^{2} \sin \left[\theta_{i}(t)+\psi\right]+\xi_{i}(t) \label{eq:ap_7},
\end{eqnarray}
such that $\langle \xi_i (t)\rangle=0$  and $\langle \xi_i (t) \xi_j (t^\prime )\rangle =2\gamma k_B T\delta_{ij} \delta(t-t^\prime )$,  where $\gamma$ is a damping constant. In this case, the oscillator probability density $\rho$ must satisfy the nonlinear Fokker-Planck equation \cite{43}
\begin{eqnarray}
\frac{\partial \rho}{\partial t}&=&D \frac{\partial^{2} \rho}{\partial \theta^{2}}-\frac{\partial(v \rho)}{\partial \theta}, \nonumber \\
v&=&\omega+K r^{2} \sin \left[\theta_{i}(t)+\psi\right], \label{eq:ap_8}
\end{eqnarray}
where $v$ is the drift velocity and with $D(\Omega)=\gamma k_B T\Omega$ is a diffusion coefficient. The solution to these coupled equations yields 
\begin{eqnarray}
r=\sqrt{1-\frac{K_{c}(\delta, T)}{K r}} \label{eq:ap_9},
\end{eqnarray}
and we see that the only difference to equation \eqref{eq:ap_6} is in the critical coupling, which now reads
\begin{eqnarray}
\frac{2}{K_{c}(\delta, T)}=\int_{-\infty}^{+\infty} d \Omega \frac{g(D(\Omega)+\omega)}{\Omega^{2}+1}. \label{eq:ap_10}
\end{eqnarray}
One important conclusion is that by mapping the Heisenberg equations of motion of each lateral phonon community controlled by $b_L$ and $b_R$ into a Kuramoto equation of phase oscillators, we are able to connect the Kuramoto order parameter $r$ with the anharmonicity introduced in the cluster, $K$, that controls the exchange of independent vibrations and locked, triatomic molecular vibrations related to the presence of the Sr atom in the dielectric layer. The collective behavior of IQTPs is caused by the synchronization of the phonon communities present in the $\mathrm{Cu-O_{ap}}$ pairs. Furthermore, the overall effect of temperature (dissipation through thermal noise) is to increase the critical coupling $K_c$ of the ground state, making it harder for the oscillators to synchronize. The temperature increases the spread in the Lorenztian probability distribution for the oscillators and this leads to an increase of the critical anharmonicity for synchronization. Also, temperature introduced in connection to dissipation enhances decoherence, which leads to an increase of the critical coupling for synchronization. Elevating the temperature does not alter the character of the synchronization transition, it remains first order.

\section{The nonlinear multimodal Bogoliubov transformation } \label{sec:appendix B}

The correlated, anharmonic phonon Hamiltonian \eqref{eq: el}-\eqref{eq:sr} of the main text can be diagonalized by means of a multimodal, nonlinear Bogoliubov transformation. To this end we introduce new phonons $B_r$,$B_r^\dagger$ that connect to the original ones through 
\begin{eqnarray}
b_{L} & \;=\; & \sum_{r}\left(u_{L,r}^{*}\beta_{r}B_{r}+v_{L,r}\beta_{r}^{\dagger}B_{r}^{\dagger}\right),\; \;\; \; \; r = R, C\label{eq:apB_1}\\
b_{R} & = & \sum_{r}\left(u_{R,r}\beta_{r}^{\dagger}B_{r}^{\dagger}+v_{R,r}^{*}\beta_{r}B_{r}\right), \;\;\;\;\; r = L,C
\label{eq:apB_2}
\end{eqnarray}
It is important to emphasize that, while the original $b_L$,$b_R$ phonons are restricted to the L,R sites only, here $r$ is allowed to run over the other positions: $R, C$ for the left phonons and $L, C$ for the right ones. This is what provides the transformation with a multimodal character. The nonlinear aspect results from the presence of quadratic terms such as $\beta_{r} B_{r}.$

Unitarity of the Bogoliubov transformation guarantees that the original commutation relations are preserved. We started with L and R phonons only, satisfying the commutation relations 
\begin{eqnarray}
\left[b_{r},b_{r^\prime}^{\dagger}\right]=\delta_{r,r^\prime} ,\qquad r,r^\prime=L,R
\end{eqnarray}

Let us consider the simpler case where $r=r^\prime$, in which case $[b_r,b_r^\dagger ]=1$, and impose the bosonic commutation relations 
\begin{eqnarray}
\left[B_{r},B_{r^{\prime}}^{\dagger}\right] & = & \delta_{r,r^{\prime}},\\
\left[\beta_{r},\beta_{r^{\prime}}^{\dagger}\right] & = & \delta_{r,r^{\prime}},
\label{eq:apB_3}
\end{eqnarray}
and expectation values of all phonon operators 
\begin{eqnarray}
n_{\beta}\left(r\right)& = & \left\langle \beta_{r}^{\dagger}\beta_{r}\right\rangle ,\\
n_{B}\left(r\right) & = & \left\langle B_{r}^{\dagger}B_{r}\right\rangle .
\label{eq:apB_4}
\end{eqnarray}
Using $b_L$ and $b_L^\dagger$, for example, the unitarity of the transformation reduces to a sum rule 
\begin{eqnarray}
\left[b_{L},b_{L}^{\dagger}\right] & = & 
\sum_{r}\left(|u_{L,r}|^2 - |v_{L,r}|^2\right)\left(1 + n_B(r) + n_\beta(r)\right) = 1.\nonumber
\label{eq:apB_5}
\end{eqnarray}
with a similar result for the right phonons, taking $L\rightarrow R$. After performing the Bogoliubov transformation, substituting equations (\ref{eq:apB_1}) and (\ref{eq:apB_2}) in the Hamiltonian, one has to eliminate the nonlinear, anharmonic interaction by setting to zero the equation for the off-diagonal Bogoliubov coefficients. Since we are interested in studying what happens in the central, planar site of the cluster, after the synchronization transition, we restrict the transformation to the coefficients $u_{r,C}$ and $v_{r,C}$, neglecting contributions coming from combinations of left/right coefficients, such as $u_{L,L}$ or $v_{R, L}$, for example. Therefore, the equation we have to eliminate from the off-diagonal elements takes the form
\begin{eqnarray}
\sum_{r=L,R} \hbar\omega u_{rC}v_{rC}+ \frac{K}{2}\sqrt{n_\beta (C)} \left(u_{rC}u_{-r,C} + v_{rC}v_{-r,C}\right)= 0.\nonumber
\label{eq:apB_6}
\end{eqnarray}
At this point, a few approximations are in order. First, we use that $v_{-r,C}=v_{r,C}^*$ and the same for the other Bogoliubov coefficients. From the numerical results of the exact diagonalization, we know that the Sr-phonon occupation in the central position becomes greater than the lateral ones, thus we shall restrict our subsequent analysis to the strong anharmonicity case, $K\gg K_c$, and approximate  
\begin{eqnarray}
\left\langle\beta_{r}^{\dagger} \beta_{r^{\prime}}^{\dagger}\right\rangle \approx\left\langle\beta_{r}^{\dagger}\right\rangle\left\langle\beta_{r^{\prime}}^{\dagger}\right\rangle=\sqrt{n_{\beta}(C)} \sqrt{n_{\beta}(C)} \delta_{r, C} \delta_{r^{\prime}, C}, \nonumber
\end{eqnarray}
which is basically the same approximation we have already used when discussing Kuramoto’s mean field equation. Using these approximations, we are left with
\begin{eqnarray}
\sum_r  \left[ \hbar\omega u_{r,C}v_{r,C} + \frac{K}{2}\sqrt{n_\beta(C)}\left(|v_{r,C}|^2 + |u_{r,C}|^2 \right) \right] = 0.\nonumber
\end{eqnarray}
Solving for the expression inside square brackets we obtain the Bogoliubov coefficients 
\begin{eqnarray}
u_{r,C} = \frac{1}{\sqrt{A}}\left(\frac{\omega + \gamma}{2\gamma}\right)^{1/2},\nonumber\\
v_{r,C} = \frac{1}{\sqrt{A}}\left(\frac{\omega - \gamma}{2\gamma}\right)^{1/2},
\end{eqnarray}
where $\hbar\gamma = \sqrt{(\hbar\omega)^2 - K^2n_\beta(C)}$ and $A=1+n_\beta (C)+n_B (C)$. We see that in this limit the coefficients are actually indepedent of $r$, enabling us to use only $u_C$ and $v_C$. Now we are ready to write down the total diagonal phonon Hamiltonian in terms of the new Bogoliubov phonons by substitution of the transformation inside the diagonal, harmonic term of the Hamiltonian in Eq. (\ref{eq: ph}) and retaining only the central contributions. This yields
\begin{eqnarray}
H_{d}^{\prime}= \hbar F(K) B_{C}^{\dagger} B_{C} ,\label{eq:apb_8}
\end{eqnarray}
where the natural frequencies $\hbar F$ are given by the expression 
\begin{eqnarray}
\hbar F(K) & = & \frac{\hbar\omega}{A\gamma} \left[\omega(1+2n_\beta(C)) - \gamma\right]\nonumber\\
& + & \frac{K\sqrt{n_\beta(C)}}{A\gamma} \sqrt{\omega^2 - \gamma^2}(1+2n_\beta(C)).
\end{eqnarray}
We can see that, before synchronization, $K<K_c$, $\hbar F=0$, because $n_\beta (C)=0$, which is expected when compared with the numerical results shown in the main text, {\it i.e.}, the new phonons are only present after the central site becomes activated by the synchronization of the IQTPs. We have to also consider the effect of the Bogoliubov transformation in the original electron-phonon coupling of the Hamiltonian in  (Eq. \ref{eq: elph}). In terms of the new phonon operators we have for $b_L + b_L^\dagger$, as an example 
\begin{eqnarray}
\left\{ \sum_r \left(u^*_{L,r}\beta_r B_r + v_{L,r}\beta^\dagger_r B^\dagger_r\right) + \sum_r \left(u_{L,r}\beta^\dagger_r B^\dagger_r + v^*_{L,r}\beta_r B_r\right)  \right\},\nonumber
\label{eq:apB_10}
\end{eqnarray}
where by performing the same procedure we used for the diagonal, harmonic terms $b_L^\dagger b_L$ and $b^\dagger_R b_R$, we write the transformed electron-phonon Hamiltonian as 
\begin{eqnarray}
H_{e l-p h}^{'}=2\lambda n_{ap}\sqrt{n_{\beta}(C)}\left(u_C + v_C\right)\left(B_C + B_C^\dagger\right),\nonumber
\end{eqnarray}
where we used the numerical result from Fig. \ref{Fig: 2}b) that $n_{el}(L) = n_{el}(R)$ for all $K$ to simplify the notation to $n_{ap}$, the electronic occupation in the apical positions of the cluster. Finally, in terms of the Bogoliubov coefficients and together with diagonal part \eqref{eq:apb_8}, the newly transformed Hamiltonian with a novel central electron-phonon coupling is written as $H^\prime =H_d^\prime +H_{el-ph}^{'}$
\begin{eqnarray}
H^{\prime}=\hbar F(K) B_{C}^{\dagger} B_{C}+ \lambda^{\prime} n_{ap}\left(B_{C}+B_{C}^{\dagger}\right),\nonumber
\end{eqnarray}
where the new coupling depends on the central Sr-phonon related occupation $n_\beta (C)$ as
\begin{eqnarray}
\lambda^\prime = \frac{2\lambda \sqrt{n_\beta (C)}}{\sqrt{2A\gamma}}\left(\sqrt{\omega + \gamma} + \sqrt{\omega - \gamma}\right).\label{eq:apB_11}
\end{eqnarray}

To summarize, by separating the $\mathrm{O_{ap}-Sr-O_{ap}}$ triatomic molecule and using a nonlinear, multi-modal Bogoliubov transformation, we started with two oscillator communities associated with $b_L,b^\dagger_L$ and $b_R,b^\dagger_R$ phonon modes, but we ended up with three oscillator communities, with the addition of the modes associated with $B_C,B_C^\dagger$. This introduces a novel central phonon, located in the plane and related to the planar oxygen oscillations, together with the original left and right oscillations of the apical oxygens. Furthermore, those novel planar vibrational modes couple to the electronic degrees of freedom with a renormalized electron-phonon coupling given by equation \eqref{eq:apB_11}, which gives rise to a new IQTP mode, now located in the copper oxide plane. To see how exactly the new phonon, $B_C$, couples to the central oxygen atom, $\mathrm{O_{pl}}$, let us recall that within our 6-atom cluster one has a single excess hole for all oxygen atoms
\begin{equation}
2n_{ap}+n_{C}=1 \rightarrow n_{ap} = \frac{1 - n_C}{2},
\end{equation}
where $n_C$ is the electronic occupation of the planar $\mathrm{O_{pl}}$ at the center of the cluster. So, after applying the non-linear Bogoliubov transformation we rewrite the central contribution of the electron-phonon transformed Hamiltonian in a way that the new phonon $B_C$ couples to the central oxygen atom, via its occupation, after the synchronization transition, when $n_\beta (C)>0$. Therefore, after a multimodal, non-linear Bogoliubov transformation, not only a newly phonon mode, associated to the planar oxygen displacements, appears, but it also couples to the excess charge pumped to the central site, giving rise to an extension of the IQTP’s to the plane and allowing the description of the problem to be summarized in the triple-well potential interpretation.


\end{document}